\documentclass[aip,twocolumn,numerical,superscriptaddress,nofootinbib,showkeys]{revtex4}
\usepackage{amssymb}
\usepackage{multirow}
\usepackage{array}
\usepackage{enumerate}
\usepackage{footnote}
\usepackage{graphicx}

\begin{document}

\title{Characterization of the QUartz Photon Intensifying Detector (QUPID)\\for Noble Liquid Detectors}

\author{A.~Teymourian}
\altaffiliation{Corresponding Author, Tel: (310) 825-1902,\\email: artintey@physics.ucla.edu}
\affiliation{Department of Physics and Astronomy, University of California, Los Angeles,\\475 Portola Plaza, Los Angles, CA 90095, USA}

\author{D.~Aharoni}
\affiliation{Department of Physics and Astronomy, University of California, Los Angeles,\\475 Portola Plaza, Los Angles, CA 90095, USA}
\author{L.~Baudis}
\affiliation{Physics Institute, University of Z\"urich, Winterthurerstrasse 190, CH-8057, Z\"urich, Switzerland}
\author{P.~Beltrame}
\affiliation{Department of Physics and Astronomy, University of California, Los Angeles,\\475 Portola Plaza, Los Angles, CA 90095, USA}
\author{E.~Brown}
\altaffiliation{Current Address: Institut f\"ur Kerphysik, Westf\"alische Wilhelms-Universit\"at M\"unster, 48149 M\"unster, Germany}
\affiliation{Department of Physics and Astronomy, University of California, Los Angeles,\\475 Portola Plaza, Los Angles, CA 90095, USA}
\author{D.~Cline}
\affiliation{Department of Physics and Astronomy, University of California, Los Angeles,\\475 Portola Plaza, Los Angles, CA 90095, USA}
\author{A.D.~Ferella}
\affiliation{Physics Institute, University of Z\"urich, Winterthurerstrasse 190, CH-8057, Z\"urich, Switzerland}
\author{A.~Fukasawa}
\affiliation{Electron Tube Division, Hamamatsu Photonics K.K., 314-5 Shimokanzo, Iwata City 438-0193, Shizuoka, Japan}
\author{C.W.~Lam}
\affiliation{Department of Physics and Astronomy, University of California, Los Angeles,\\475 Portola Plaza, Los Angles, CA 90095, USA}
\author{T.~Lim}
\affiliation{Department of Physics and Astronomy, University of California, Los Angeles,\\475 Portola Plaza, Los Angles, CA 90095, USA}
\author{K.~Lung}
\affiliation{Department of Physics and Astronomy, University of California, Los Angeles,\\475 Portola Plaza, Los Angles, CA 90095, USA}
\author{Y.~Meng}
\affiliation{Department of Physics and Astronomy, University of California, Los Angeles,\\475 Portola Plaza, Los Angles, CA 90095, USA}
\author{S.~Muramatsu}
\affiliation{Electron Tube Division, Hamamatsu Photonics K.K., 314-5 Shimokanzo, Iwata City 438-0193, Shizuoka, Japan}
\author{E.~Pantic}
\affiliation{Department of Physics and Astronomy, University of California, Los Angeles,\\475 Portola Plaza, Los Angles, CA 90095, USA}
\author{M.~Suyama}
\affiliation{Electron Tube Division, Hamamatsu Photonics K.K., 314-5 Shimokanzo, Iwata City 438-0193, Shizuoka, Japan}
\author{H.~Wang}
\affiliation{Department of Physics and Astronomy, University of California, Los Angeles,\\475 Portola Plaza, Los Angles, CA 90095, USA}
\author{K.~Arisaka}
\affiliation{Department of Physics and Astronomy, University of California, Los Angeles,\\475 Portola Plaza, Los Angles, CA 90095, USA}

\begin{abstract}
\begin{center}
  \line(1,0){400}
\end{center}

Dark Matter and Double Beta Decay experiments require extremely low radioactivity within the detector materials. For this purpose, the University of California, Los Angeles and Hamamatsu Photonics have developed the QUartz Photon Intensifying Detector (\textsc{Qupid}), an ultra-low background photodetector based on the Hybrid Avalanche Photo Diode (HAPD) and entirely made of ultraclean synthetic fused silica. In this work we present the basic concept of the \textsc{Qupid} and the testing measurements on \textsc{Qupid}s from the first production line.

Screening of radioactivity at the Gator facility in the Laboratori Nazionali del Gran Sasso has shown that the \textsc{Qupid}s safely fulfill the low radioactive contamination requirements for the next generation zero background experiments set by Monte Carlo simulations.

The quantum efficiency of the \textsc{Qupid} at room temperature is $>30$\% at the xenon scintillation wavelength. At -100$^{\circ}$~C, the \textsc{Qupid} shows a leakage current smaller than 1 nA and a global gain of $10^5$.  In these conditions, the photocathode and the anode show $>95\%$ linearity up to $1~\mu$A for the cathode and 3~mA for the anode. The photocathode and collection efficiency are uniform to 80\% over the entire surface. In parallel with single photon counting capabilities, the \textsc{Qupid}s have a good timing response: $1.8 \pm 0.1$~ns rise time, $2.5 \pm 0.2$~ns fall time, $4.20 \pm 0.05$~ns (FWHM) pulse width, and $160 \pm 30$~ps (FWHM) transit time spread.

The \textsc{Qupid}s have also been tested in a liquid xenon environment, and scintillation light from $^{57}$Co and $^{210}$Po radioactive sources were observed.

\begin{center}
 \line(1,0){400}
\end{center}

\end{abstract}

\maketitle

\section{Introduction}
\label{sec:introduction}

There is overwhelming indirect evidence that Dark Matter (DM) accounts for $\sim\:$25\% of the mass-energy of the universe. One of the most accredited theories predicts that Weakly Interacting Massive Particles (WIMPs) constitute the dark matter halos that permeate galaxies~\cite{DMint}. WIMPs are expected to undergo elastic collisions with noble liquid nuclei producing low energy deposits~\cite{DMdet}. Some of the noble liquid experiments searching for WIMP interactions are XENON, ZEPLIN, LUX, DAMA/LXe and XMASS~\cite{Xenon10,Xenon100,Aln,LUX,DAMA,XMASS} employing liquid xenon, and WARP, DEAP/CLEAN and \textsc{DarkSide}, using liquid argon~\cite{WARP,DEAPCLEAN,DarkSide}.  One promising approach for noble liquid experiments is the double phase Time Projection Chamber (TPC).  Here, a WIMP deposits energy in the noble liquid creating a primary scintillation light (S1).  The energy deposit also ionizes atoms, and the freed electrons are drifted to a gas phase where they undergo a secondary proportional scintillation (S2).
\begin{figure*}
  \begin{center}
    \includegraphics[height=55mm]{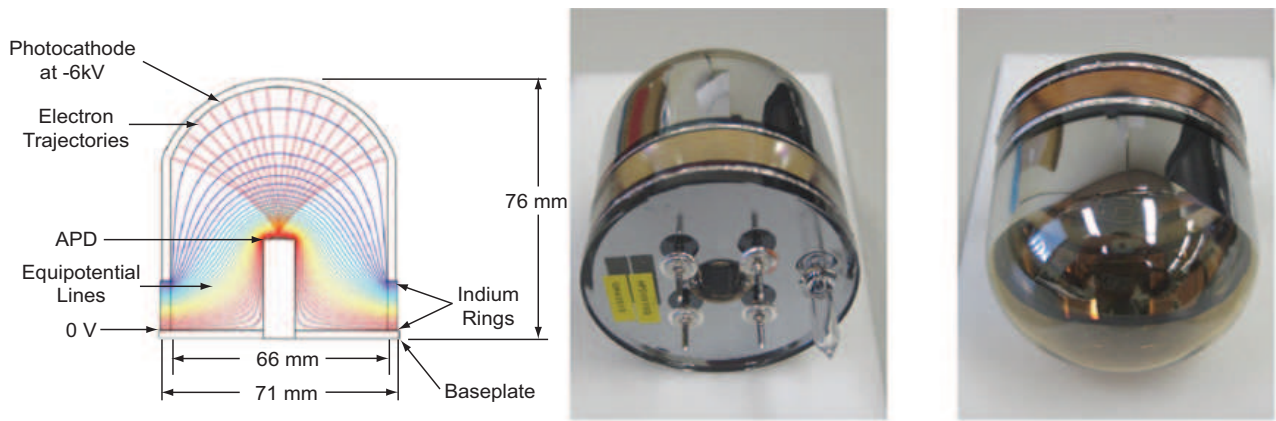}
  \end{center}
  \caption{
    \label{fig:Qupid_photo} 
     Drawing of the \textsc{Qupid} showing the electric field and the electrons trajectory simulations, {\it left} panel. It can be seen that the photoelectrons are focused onto the APD. Back and front views of the \textsc{Qupid} are shown in the {\it center} and {\it right} panels, respectively. Two indium rings, one used to provide -6 kV to the photocathode and the other for grounding, can be seen. The same rings hold the quartz cylinder, quartz ring, and baseplate together. On the baseplate, one can see four pins (two of them are connected to the APD cathode and anode, while the other two are used only during production). The glass pipe is used for pumping a vacuum during production.}
\end{figure*}

	 Natural xenon contains the isotope $^{136}$Xe, which is predicted to undergo double beta decay, normally accompanied by the production of two neutrinos ($2\nu\beta\beta$). The detection of a peak in the summed energy of two electrons at 2.458 MeV would represent the unequivocal signature of neutrinoless double beta decay ($0\nu\beta\beta$), proving that the neutrino is its own antiparticle (a Majorana particle)~\cite{QVal,DBD}. Experiments using $^{136}$Xe such as EXO, NEXT, and KamLAND-Zen~\cite{EXO,EXO2,Next,Kamland} are underway.\\
\indent Presently, the most sensitive dark matter detector using noble liquids has an effective target mass on the order of 100 kg with an expected background rate of $\sim\:10^{-2}$~events/kg/day/keV\footnote{keVee and keVr are used to distinguish between electron recoil and nuclear recoil interactions. Throughout this paper, we will use keV in place of keVee.} in the region of interest~\cite{Xenon100BG}.  Next generation detectors, with target mass $>1$ ton, will require a background rate $< 10^{-4}$~events/kg/day/keV to achieve the desired sensitivites. In order to decrease the background down to this quantity, several efforts are underway~\cite{XAX, Xenon1T}. Of particular importance is the radioactive contamination coming from the employed photodetectors, as they are the experimental components closest to the target material. Moreover, in present experiments the majority of the background rate originates from the conventional photomultiplier tubes (PMTs).\\
\indent A new photodetector, the QUartz Photon Intensifying Detector, or \textsc{Qupid}, has been designed as an ideal replacement for conventional PMTs\footnote{US Patent Pending, Application No. 20100102408, M. Suyama, A. Fukasawa, K. Arisaka, H. Wang} by the University of California, Los Angeles (UCLA) and Hamamatsu Photonics.  In this work, we present the development and tests performed on a set of seven \textsc{Qupid}s from an early production line. An explanation of the essential requirements that must be satisfied for dark matter and double beta decay detection, and the general \textsc{Qupid} concept are presented in Sec.~\ref{sec:requirements} and \ref{sec:concept} respectively. In Sec.~\ref{sec:radioactivity} we discuss the radioactive screening of the \textsc{Qupid} at the Gator screening facility at the Laboratori Nazionali del Gran Sasso (LNGS) in Italy, performed by the University of Zurich group.\\
\indent In Sec.~\ref{sec:photocathode} we examine the cathode performances and those of the anode in  Sec.~\ref{sec:anode}. Sec.~\ref{sec:shapingtiming} is devoted to single photon counting and time response tests, while in Sec.~\ref{sec:liquidxenon} we show the operation of the \textsc{Qupid} in liquid xenon\footnote{Although a modified version of the \textsc{Qupid} optimized for operating in liquid argon is in development, in this work we will concentrate on the \textsc{Qupid} for liquid xenon.}. In Sec.~\ref{sec:conclusions} we present a summary of the main achievements, proving that the \textsc{Qupid}s represent the most appropriate solution for the next generation of low background detectors.
\section{Photodetector Requirements}
\label{sec:requirements}

The photodetectors which will be employed in the next generation experiments must meet the following requirements in order to improve the performance both for reducing the background contamination and increasing sensitivity:
\begin{itemize}
\item {\it intrinsic radioactivity} significantly lower than the current generation PMTs;
\item {\it Quantum efficiency} $>30\%$ to maximize the number of photoelectrons per deposited energy;
\item high {\it gain} $>10^{5}$ so that single photoelectrons can be clearly detected above the noise;
\item good {\it timing} performances with a pulse width $<~10$~ns, to provide accurate time information.
\item good {\it collection efficiency} and {\it uniformity} along the photodetector surface;
\item a large dynamic range with superior {\it linearity}, for precise measurement of energy depositions in the region of interest: from a few keV for dark matter searches to several MeV for neutrinoless double beta decay;  
\end{itemize}
In the following sections we describe in some detail the tests performed proving the capability for the \textsc{Qupid} to satisfy these requirements.

\begin{table*}
  \centering
    \setlength{\extrarowheight}{1.5pt}
  \setlength{\tabcolsep}{10pt}

  \begin{tabular}{|@{}l|c|c|c|c|}
    \hline
    \multirow{3}{*}{Contaminant} & \multirow{3}{*}{Activity (mBq/\textsc{Qupid})} & \multicolumn{3}{c|}{Events/year in Fiducial Cut,} \\
    & & \multicolumn{3}{c|}{Target Mass (after fiducial cut), Energy [2-18] keV}\\
    \cline{3-5}
    & & 0 cm, 2.3 ton & 5 cm, 1.6 ton & 10 cm, 1.1 ton \\
    \hline
    \hline
    $^{238}$U & $<17.3$ & $<560$ & $<0.5$ & 0\\
    $^{226}$Ra & $0.3\pm 0.1$ & 23 & 0.14 & 0.01\\
    $^{232}$Th & $0.4\pm 0.2$ & 35 & 0.24 & 0.02\\
    $^{40}$K & $5.5\pm 0.6$ & 55 & 0.32 & 0.02\\
    $^{60}$Co & $<0.18$ & $<4.9$ & $<0.21$ & $<0.02$\\
    \hline
    Total & $<23.7$ & $<678$ & $<1.41$ & $<0.07$\\
    \hline
  \end{tabular}
  \caption{First column: Contaminants present in the \textsc{Qupid} divided into the active chains. Second column: Measured intrinsic radioactivity of the \textsc{Qupid}s. Remaining columns: Radioactive background from the \textsc{Qupid} in a ton-scale detector for different fiducial volume cuts in the 2-18 keV energy range. The relatively high contamination arising from $^{238}$U, of about 17 mBq per \textsc{Qupid}, does not affect the region of interest as $\gamma$-rays from this chain do not penetrate deeply inside the liquid xenon.  It can be easily cut out, down to zero events per year, by increasing the fiducial cut to 10 cm from each side.} 
  \label{tab:Xenon1TSim}
\end{table*}   
\section{Concept}
\label{sec:concept}

The \textsc{Qupid} is based on the Hybrid Avalanche Photodiode (HAPD). In HAPDs, photons hit the photocathode surface causing the emission of photoelectrons, which are accelerated onto an APD due to a high potential difference (several kV) applied between the photocathode and the APD. The kinetic energy of the electrons creates hundreds of electron-hole pairs within the APD. The electrons and holes are then separated and accelerated by of the high bias voltage on the APD, and undergoing an avalanche effect~\cite{Suyam,Nakam,Magic}.\\
\indent On the left of Fig.~\ref{fig:Qupid_photo}, a drawing of the \textsc{Qupid} is presented, showing a simulation of the electric field equipotential lines and of the photoelectron trajectories from the photocathode onto the APD.  Fig.~\ref{fig:Qupid_photo} center shows the \textsc{Qupid} seen from the baseplate, and on the right from the hemispherical photocathode.\\
\indent The \textsc{Qupid} is made of a cylindrical quartz tube with a hemispherical photocathode window, a baseplate, and an intermediate quartz ring.  The quartz cylinder, ring, and baseplate are bonded together using indium (see Fig.~\ref{fig:Qupid_photo} center and right). The outer diameter of the cylinder is 71 mm, with the hemispherical photocathode window having a radius of 37 mm. The photocathode has an effective diameter of 64 mm (for vertical incident photons). The inner part of the cylinder is coated with aluminum and the hemispherical cap is coated with a photocathode material.  The baseplate on the opposite end supports a solid cylindrical quartz pillar with a 3 mm APD at the top. The APD (with 11 pF of capacitance) has been specifically developed and manufactured by Hamamatsu Photonics for use in the \textsc{Qupid}s.\\
\indent A negative high voltage, up to -6 kV, is applied to the photocathode through the indium sealing ring, while ground level is maintained on the baseplate and APD from the second indium ring. As in common HAPDs, the high potential difference creates an electric field which focuses the photoelectrons ejected from the photocathode onto the APD. The \textsc{Qupid} design has been optimized such that the photoelectron focusing is independent of the level of voltage applied to the photocathode.
\section{Radioactivity}
\label{sec:radioactivity}

The radioactivity of the \textsc{Qupid}s has been measured in the Gator screening facility, operated by the University of Zurich at LNGS. The facility consists of a high-purity, p-type coaxial germanium (HPGe) detector with a 2.2~kg  sensitive mass, operated in an ultra-low background shield continuously flushed with boil-off nitrogen gas to suppress radon diffusion. With an integral background rate of 0.16 events/min in the 100-2700 keV region, Gator is one of the world's most sensitive Ge spectrometers~\cite{Gator}.  In this work we present the results of the screening of two batches of five \textsc{Qupid}s.\\
\indent To determine the specific activities for the $^{238}$U and $^{232}$Th chains, as well as for $^{60}$Co and $^{40}$K, the  most prominent $\gamma$-lines of the respective decays are analyzed using the efficiencies determined by a detailed Monte Carlo simulation of the detector, shield, and \textsc{Qupid} samples.  The latest background run of Gator had been taken for a duration of two months prior to the \textsc{Qupid} screening.  In the case in which no events were detected above the background, upper limits on the specific activities were calculated according to the method proposed in ~\cite{Hurtgen}.\\
\indent The radioactivities for each \textsc{Qupid} are $<17.3$ mBq/\textsc{Qupid} for $^{238}$U, $0.3\pm 0.1$ mBq/\textsc{Qupid} for $^{232}$Th, $0.4\pm 0.2$ mBq/\textsc{Qupid} for $^{40}$K and  $<0.18$ mBq/\textsc{Qupid} for $^{60}$Co. These results\footnote{Although $^{226}$Ra belongs to the $^{238}$U chain with a half-life of 1600 years, it bonds easily with other electronegative elements and can be absorbed in strong thermal and/or chemical processes, generating a break in the equilibrium of the chain. For this reason, the two parts of the chain (i.e.  $^{238}$U and  $^{226}$Ra) have been treated separately in the analysis.} are reported in the first column of Table~\ref{tab:Xenon1TSim}.  The radioactivity levels of standard PMTs used in current dark matter detectors are available in \cite{MaterialScreening}.  Although in many cases only upper limits are available, the radioactivity of the \textsc{Qupid}s are consistently comparable or lower than conventional PMTs when considering the effective areas of the PMTs and \textsc{Qupid}s. 

\begin{figure*}
  \begin{center}
    \includegraphics[ height=60mm ]{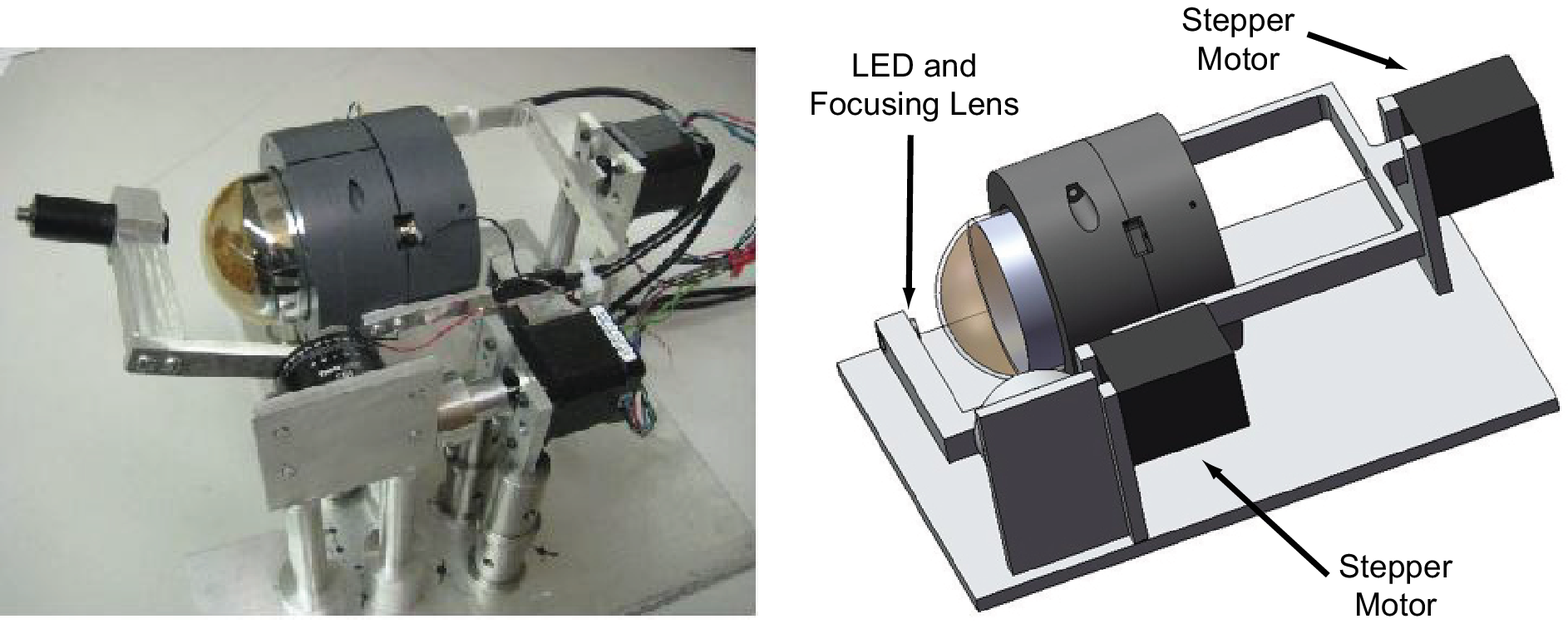}
  \end{center}
  \caption{
    \label{fig:UniformitySystem} 
    On the {\it left}, photograph of the \textsc{Qupid} setup for uniformity measurements.  On the {\it right}, its 3D rendering. In this system, the \textsc{Qupid} is rotated along the $\phi$ axis, and the LED scans along the $\theta$ axis.}
\end{figure*}
\begin{figure}
  \begin{center}
    \includegraphics[height=65mm ]{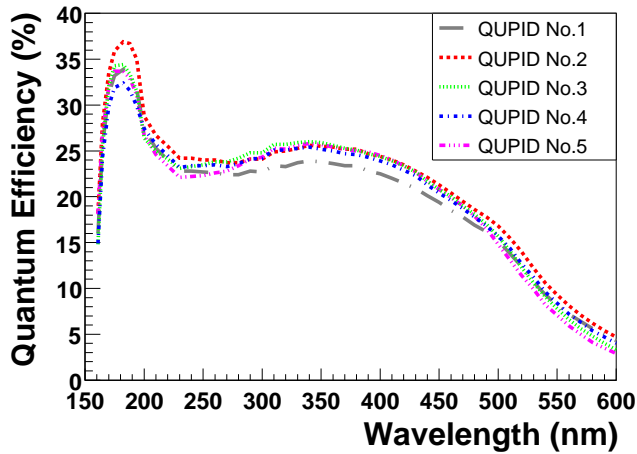}
  \end{center}
  \caption{
    \label{fig:QupidQE} 
    Quantum efficiency measured for various \textsc{Qupid}s, optimized for liquid xenon operation, with the maximum at 178 nm being $34\pm 2\%$. The numbering of the \textsc{Qupid}s are arbitrary and do not follow the production numbers.}
\end{figure}
\begin{figure*}
  \begin{center}
    \includegraphics[ height=75mm ]{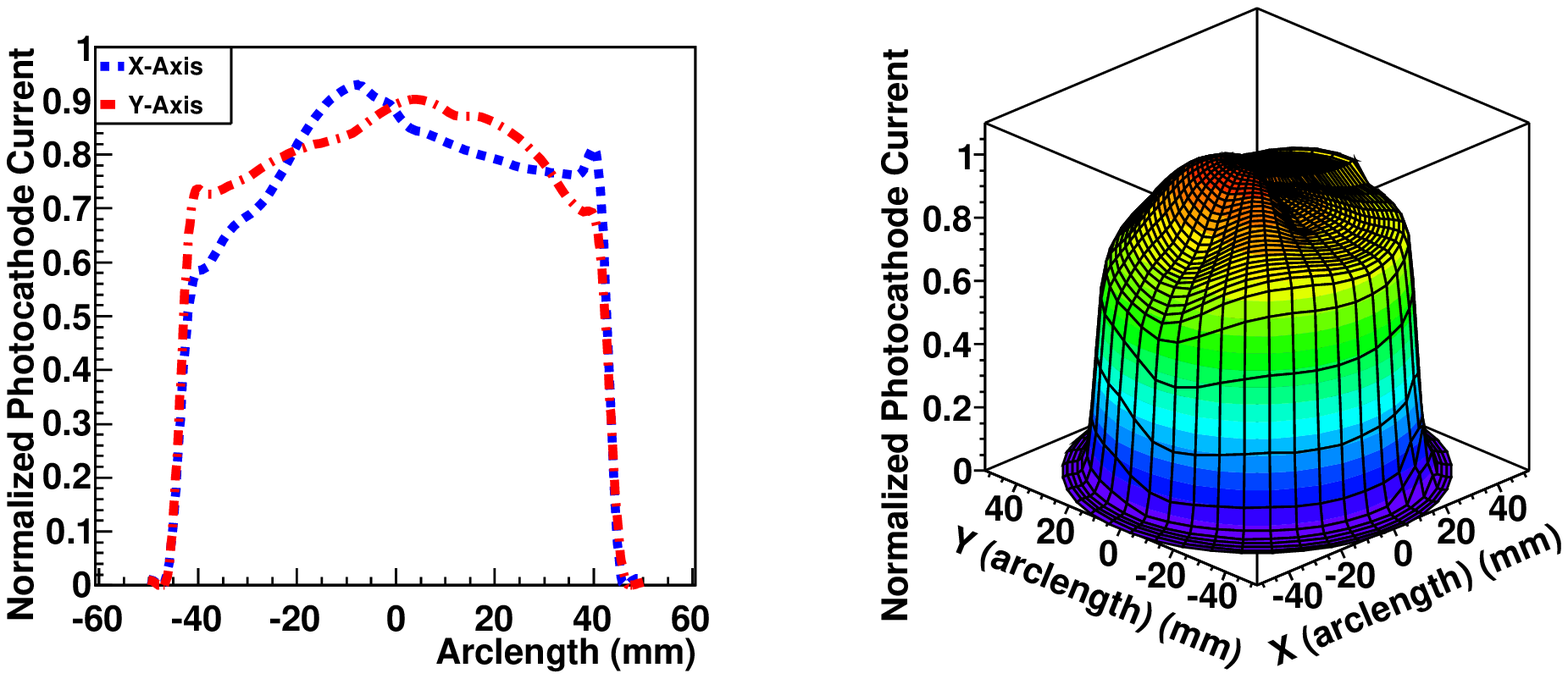}
  \end{center}
\caption{
  \label{fig:CathodeLin} 
Photocathode uniformity results for \textsc{Qupid} (No.7) showing X and Y slices (on the {\it left}), and a 3D plot (on the {\it right}). The \textsc{Qupid} is uniform to $\sim\:80\%$ across the face.}
\end{figure*}

To verify whether the background coming from the \textsc{Qupid}s match the requirements of future detectors, a preliminary design of a ton-scale liquid xenon detector has been studied using the \texttt{Geant4} Monte Carlo software package~\cite{Geant4}. The simulation considers ({\it i}) a liquid xenon TPC, with a diameter of 1 m and height of 1~m (corresponding to a total mass of 2.3 ton), ({\it ii}) two arrays of 121 \textsc{Qupid}s each placed at the top and bottom of the TPC, ({\it iii}) all the main detector materials. To estimate the background level arising from the \textsc{Qupid}s we have implemented the code assuming the radioactive contamination from the Gator data.\\
\indent In the simulation we have considered the standard analysis cuts used by the XENON collaboration~\cite{Xenon100,XenonConf}, with a conservative assumption of 99\% rejection power on the electromagnetic background. The study has been repeated for different fiducial volumes, that is, cutting the top, bottom and lateral face of the liquid xenon and considering only the inner cylindrical volume as target material and region of interest for any energy deposit. In Table~\ref{tab:Xenon1TSim} the results of the Monte Carlo simulation for fiducial volume cuts of 0, 5, and 10 cm are presented, along with the results of the screening for each chain.  The indium used in the construction of the \textsc{Qupid}s is known to undergo $\beta$-decay, and this radioactivity was also included in the Monte Carlo simulation.  Due to the short attenuation length of $\beta$-particles in liquid xenon, no energy deposits were observed in the inner volumes of the target.\\
\indent In a 1 m $\times 1$ m liquid xenon detector with 10 cm fiducial volume cuts (corresponding to a target mass of 1.1 ton) 242 \textsc{Qupid}s would give a total background rate $<0.07$~events/year, in the energy range between 2 and 18 keV. This result, unachievable using the standard PMTs, would perfectly satisfy the requirements of the next generation dark matter experiments~\cite{Xenon1T,XAX}.  

\section{Photocathode}
\label{sec:photocathode}
\subsection{Quantum Efficiency}
\label{subsec:quantumeff}

The photocathode used in the \textsc{Qupid} has been specifically developed~\cite{Photocathode} by Hamamatsu Photonics to achieve the highest quantum efficiency (QE) for 178 nm wavelength photons, corresponding to the xenon scintillation light.\\
\indent The QE has been measured at room temperature by comparing the response of the \textsc{Qupid} to a standard PMT, calibrated by means of a NIST standard UV sensitive photodiode~\cite{PMTBook}. Fig.~\ref{fig:QupidQE} shows the QE results for different \textsc{Qupid}s measured at Hamamatsu. All the tested \textsc{Qupid}s show a maximum QE $>30\%$ around 178~nm. The sharp cutoff at 170 nm is due to absorption by the quartz window.\\
\indent A photocathode version optimized for operation in liquid argon is under development. The quartz window of the \textsc{Qupid} is opaque to argon scintillation light, and thus a wavelength shifting (WLS) material\footnote{A possible WLS material in consideration is Tetraphenyl Butadiene (TPB).} must be used to shift the scintillation light to $\sim\:400$~nm.  The photocathode used for liquid argon detectors will then have the highest QE around visible light.
\begin{figure}
  \begin{center}
    \includegraphics[ height=45mm ]{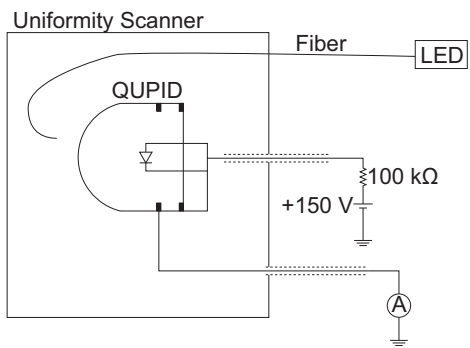}
  \end{center}
\caption{
  \label{fig:CathodeUnif} 
  Testing setup for the photocathode uniformity.  150 V were applied to the grounding ring, APD anode, and APD cathode, while the current from the photocathode was read out with a picoammeter.}
\end{figure}
\subsection{Uniformity}
\label{subsec:cathodeuniformity}

The uniformity of the \textsc{Qupid} was measured at room temperature by focusing a LED onto the photocathode and scanning over the entire face. The LED was powered by a DC power supply and provided a spot of 1 mm focused on the spherical surface of the \textsc{Qupid}, while two stepper motors controlled independently: 
\begin{itemize}
\item[({\it i})] the location of the LED along the $\theta$ axis (i.e. movement from the top of the photocathode towards the indium rings, in steps of $1^{\circ}$);
\item[({\it ii})] the position of the \textsc{Qupid} along the $\phi$ axis (i.e. movement of the \textsc{Qupid} around its main rotation axis, in steps of $10^{\circ}$).
\end{itemize}
A photo and a drawing of the setup are shown on the left and right of Fig.~\ref{fig:UniformitySystem} respectively.\\
\indent The uniformity has been tested by measuring the current from the photocathode by means of a picoammeter while applying 150 V to the grounding ring, anode, and cathode of the APD all shorted together.  This maintains an electric field that attracts the photoelectrons ejected from the photocathode.  In this scenario, the APD is not biased since there is no potential difference across the anode and cathode of the APD, and thus the voltage is only to attract the ejected photoelectrons.  A schematic of the readout for this measurement can be seen in Fig.~\ref{fig:CathodeUnif} while the results of the measurement are shown in Fig.~\ref{fig:CathodeLin}. The photocathode is uniform to about $80\%$ across its face.

\begin{figure}
  \begin{center}
    \includegraphics[height=42mm ]{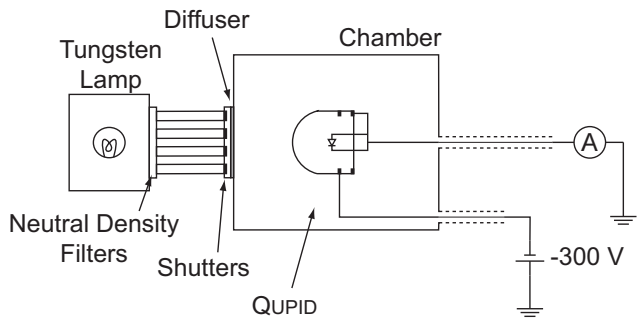}
  \end{center}
  \caption{
    \label{fig:PhotocathodeLinearityDiagram} 
    Photocathode linearity system.  A tungsten lamp, neutral density filter wheel, and diffuser, along with four shutters, control the illumination of the photocathode of the \textsc{Qupid}. The photocathode is supplied with -300 V while the APD and the grounding ring are read out by a picoammeter.}
\end{figure}
\begin{figure}
\begin{center}
\includegraphics[height=65mm ]{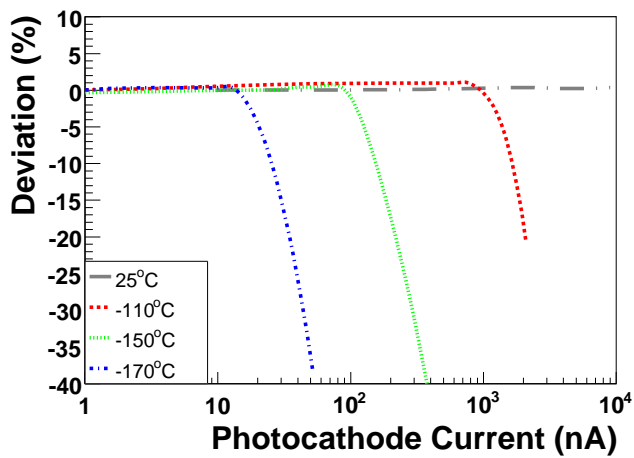}
\end{center}
\caption{
  \label{fig:BA0020Linearity} 
  Photocathode linearity versus current at various temperatures of \textsc{Qupid} (No.3). The photocathode deviates from linearity at lower currents for lower temperatures. At liquid xenon temperature, saturation occurs above 1 $\mu$A.}
\end{figure}
\begin{figure*}
  \begin{center}
    \includegraphics[ height=85mm ]{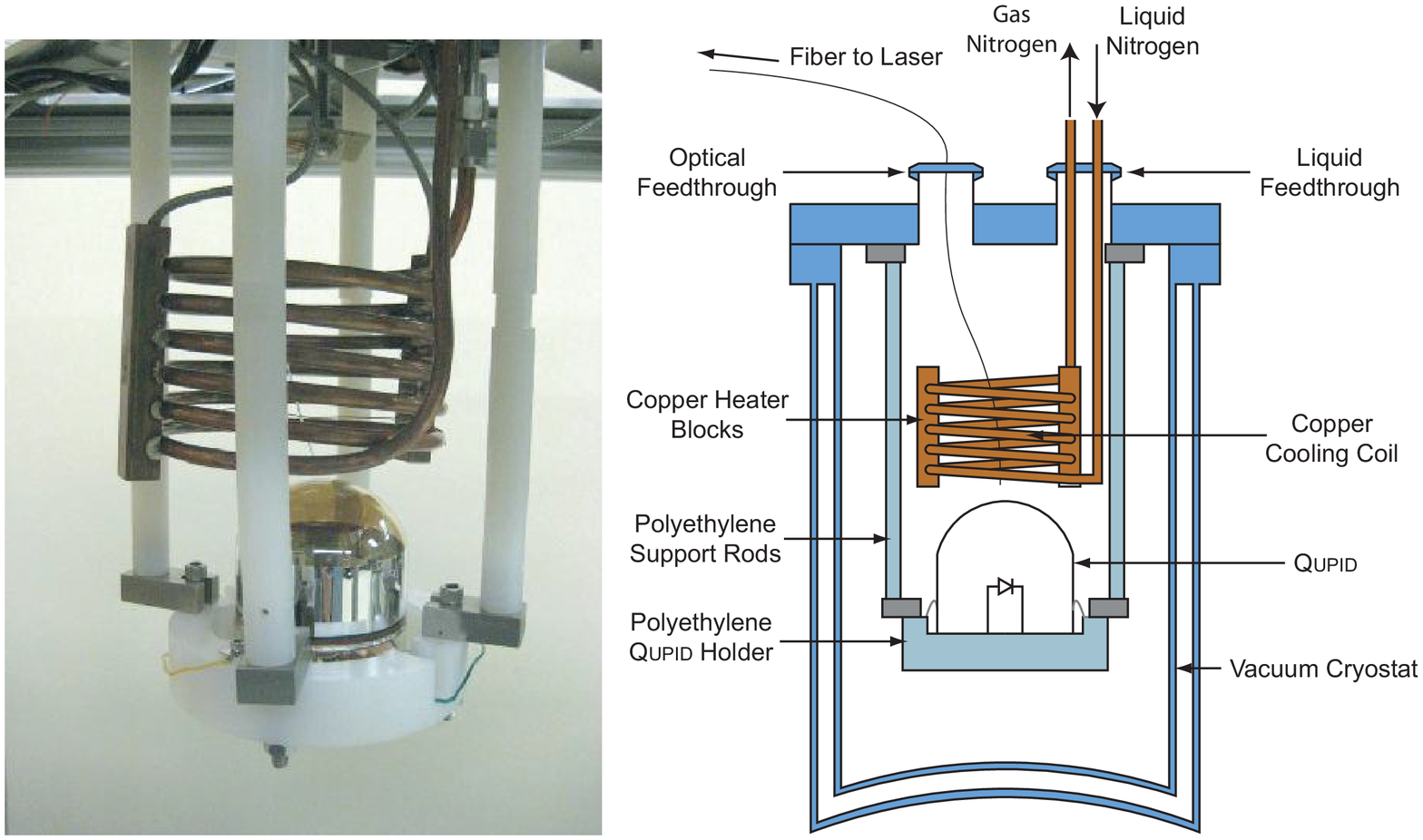}
  \end{center}
  \caption{
    \label{fig:LN2Cooling} 
    On the {\it left}, photograph of the liquid nitrogen cooling system and the \textsc{Qupid} support. On the {\it right}, diagram of the nitrogen cooling system and the \textsc{Qupid} support.}
\end{figure*}
\subsection{Linearity}
\label{subsec:cathodelinearity}

The Low Temperature Bialkali photocathode (Bialkali-LT) developed by Hamamatsu Photonics and employed in the \textsc{Qupid}s is optimized for linearity over a wide dynamic range at low operating temperatures~\cite{Photocathode}, where standard photocathodes become nonlinear as the resistivity increases.\\
\indent For neutrinoless double beta decay in $^{136}$Xe, the expected energy deposited is 2.458 MeV and the largest signals come from the S2s.  While S1 signals have a light yield of 3 photoelectrons/keV (pe/keV), corresponding to about 7,400 photoelectrons, the S2 signals have 200 times the number of photoelectrons spread out over 2~$\mu$s. The S2 signal is distributed over several photodetectors, and at most 10\% of the signal can be expected on a single \textsc{Qupid}.  This results in a maximum photocathode current of about 12~nA.  The photocathode response must then be linear up to at least this current.\\
\indent To test the linearity, the \textsc{Qupid} was uniformly illuminated by a tungsten lamp coupled with a diffuser. A set of four shutters and a neutral density filter wheel were used to vary the light intensity. The photocathode was supplied with a potential of -300 V and the APD (both the anode and the cathode) and the grounding ring were connected to a picoammeter. Fig.~\ref{fig:PhotocathodeLinearityDiagram} shows a diagram of the testing system used for the photocathode linearity.  The -300 V applied to the photocathode maintains the electric field within the \textsc{Qupid}, and the resulting photoelectron current is read with the picoammeter as the photoelectrons are accelerated to either the APD or the ground plane.  Just as in the photocathode linearity setup, the APD is not biased, and the -300 V is only used to attract the photoelectrons.\\
\indent Each of the shutters has been opened individually while keeping all the others closed, and the corresponding photocathode current was read out with the picoammeter ($I_{i}$, where $i = 1, 2, 3, 4$). The current with all the shutters opened has been measured ($I_{all}$) too. The same procedure was repeated with different filters, spanning different light intensities, thereby increasing/decreasing the photocathode current in response to the changing light intensity. In such a configuration, the deviation from linearity can be defined as 
\begin{equation}
  \Delta Lin = \frac{ I_{all} - \sum_{i=1}^4 I_i } { \sum_{i=1}^4 I_i } ~,
  \label{eq:LinEq}
\end{equation}
which in the ideal case of perfect linearity must be equal to zero for all the filtered configurations (since the sum of all photocathode currents with only one shutter open should be equal to the current with all the shutters open).\\
\indent Fig.~\ref{fig:BA0020Linearity} shows $\Delta Lin$ versus the photocathode current for different temperatures. At $-110^{\circ}C$ the linearity is well maintained up to $1~\mu$A. This value largely overcomes the 12 nA dynamic range required for neutrinoless double beta decay detection in liquid xenon.
\section{Anode}
\label{sec:anode}
\subsection{Leakage Current and Gain Measurements}
\label{subsec:leakgain}

\begin{figure}
  \begin{center}
    \includegraphics[ height=45mm ]{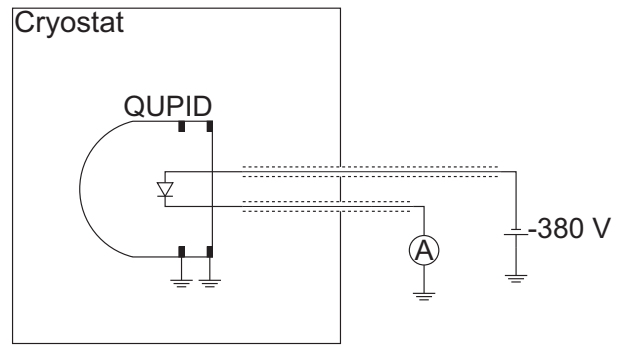}
  \end{center}
  \caption{
    \label{fig:LeakageReadout} 
    Readout system for the leakage current measurements.  The \textsc{Qupid} was placed in the cryostat with the grounding ring and photocathode grounded.  A negative bias was placed on the anode of the APD, while the leakage current was read out through the cathode of the APD with a picoammeter.}
\end{figure}

\begin{figure}
  \begin{center}
\includegraphics[ height=65mm ]{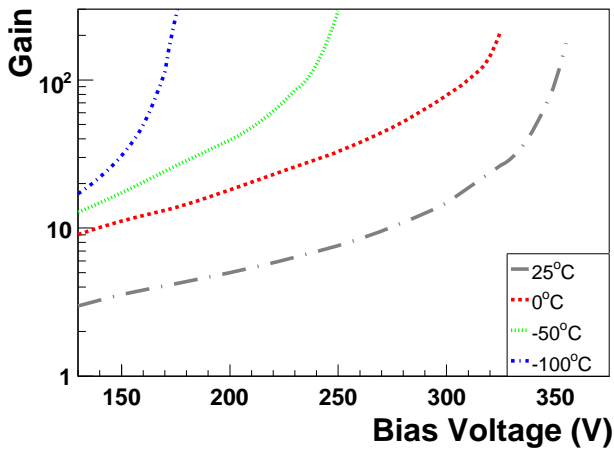}
\end{center}
\caption{
  \label{fig:BA0011_AvalancheGain} 
  Avalanche gain of \textsc{Qupid} (No.6) for various temperatures. As the temperature decreases, the avalanche gain increases for a set bias voltage. A maximum avalanche gain of 300 is seen at -100$^{\circ}$~C.}
\end{figure}

Since the main use of the \textsc{Qupid} will be in noble liquids, great care has to be taken to test its performance at cryogenic temperatures. In Fig.~\ref{fig:LN2Cooling} a picture and a drawing of the cryogenic system is shown.\\
\indent After positioning the \textsc{Qupid} in the cryostat, a vacuum is pumped using an oil free pumping station\footnote{The station (Pfieffer Vacuum Hi-Cube Eco-3 Pumping Station) consists of a diaphragm pump and a turbomolecular pump.} and dry nitrogen gas is introduced. Liquid nitrogen is then flowed through a copper coil at a constant rate to cool the surrounding nitrogen gas, cooling the \textsc{Qupid}. Two 100 W resistive heaters inserted into copper bars (soldered on the copper coil) serve as temperature stabilizers, controlled by a Proportional-Integral-Derivative (PID) controller\footnote{Omega Model CN8201 Temperature Controller} and monitored by Resistance Temperature Detectors (RTDs). At -100$^{\circ}$~C the temperature can be maintained within an accuracy of $0.1^{\circ}$ C.\\
\indent Within the cryostat, the \textsc{Qupid} is held by a polyethylene support structure anchored to the top flange of the cryostat by four polyethylene rods. A set of four stainless steel clips hold the \textsc{Qupid} to the polyethylene, which also provide connections for the photocathode voltage and the grounding ring. High voltage and coaxial feedthroughs are used to provide the photocathode voltage, bias voltage and readout. An optical fiber feedthrough is placed on the cryostat, and a fiber is pointed towards the \textsc{Qupid}. All of the readout electronics are placed outside the cryostat.\\
\indent In APDs, the leakage current increases approximately linearly with the bias voltage until breakdown is reached, above which the leakage current increases drastically.  To measure the leakage current and breakdown voltage at low temperature, the \textsc{Qupid} was placed in the cryostat, tightly insulated from external light, with both the photocathode and grounding ring grounded. A negative voltage\footnote{Stanford Research Systems Model PS350 +/- 5 kV Power Supply.} was applied to the APD anode while the cathode was connected to a picoammeter.\footnote{Keithley Model 486 Picoammeter.}  A schematic of this readout system is shown in Fig.~\ref{fig:LeakageReadout}.
\begin{figure}
  \begin{center}
    \includegraphics[ height=65mm ]{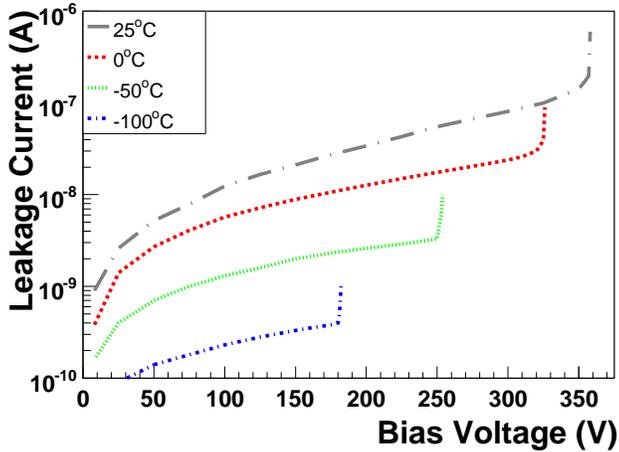}
  \end{center}
  \caption{
    \label{fig:BA0011_LeakageCurrent} 
    Leakage current versus the APD bias voltage at various temperatures of \textsc{Qupid} (No.6). As the temperature decreases, the overall leakage current decreases. Also, the breakdown voltage, recognizable from the dramatic increase of the slope of the leakage current curves, decreases with the temperature. At liquid xenon temperature, the leakage current is $<1$ nA while breakdown occurs at 180 V.}
\end{figure}
\begin{figure}
  \begin{center}
    \includegraphics[ height=65mm ]{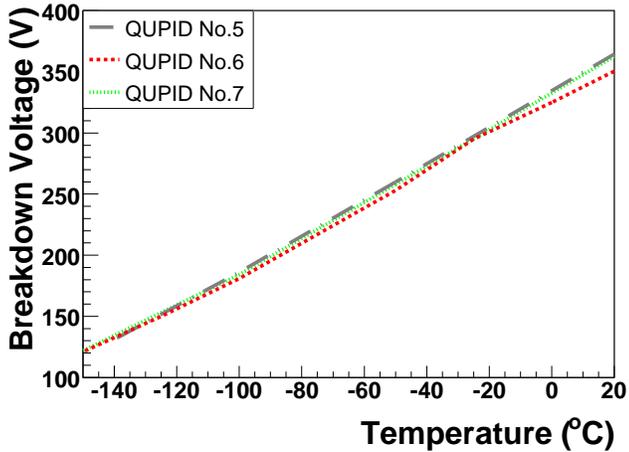}
  \end{center}
  \caption{
    \label{fig:ManyQupid_Breakdown_Temperature_Dependence} 
    Temperature dependence of the breakdown voltage for different \textsc{Qupid}s. The breakdown voltage shows a linear trend with the temperature.}
\end{figure}
\begin{figure}
  \begin{center}
    \includegraphics[ height=45mm ]{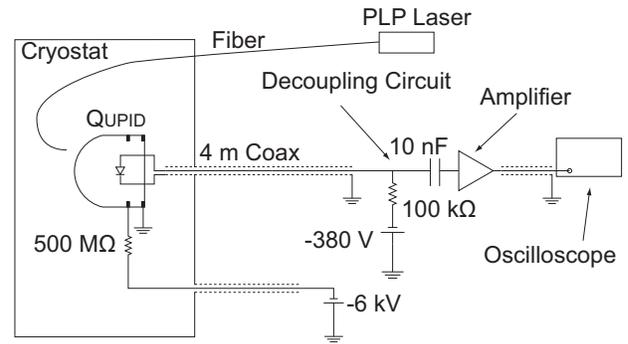}
  \end{center}
\caption{
  \label{fig:GainReadout} 
  Testing setup with the pulsed laser light used for gain and anode linearity measurement. The photocathode is supplied with high voltage and a bias voltage is connected to the APD anode through a decoupling circuit.  The output of the APD is passed through an amplifier to the readout system.}
\end{figure}
\begin{figure}
\begin{center}
\includegraphics[ height=65mm ]{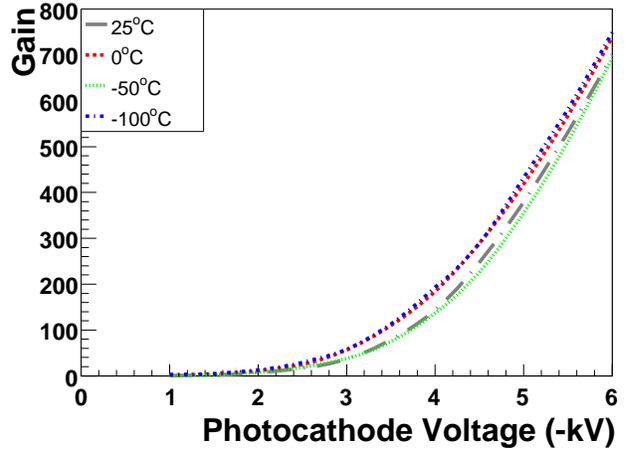}
\end{center}
\caption{
  \label{fig:BA0011_BombardmentGain} 
  Bombardment gain of \textsc{Qupid} (No.6) for various temperatures. The bombardment gain shows no temperature dependence and reaches a maximum of 750.}
\end{figure}
Fig.~\ref{fig:BA0011_LeakageCurrent} shows the leakage current curves for different temperatures as a function of the applied bias voltage. As expected, the leakage current follows a linear increase up to a breakdown voltage, where it rises dramatically. At liquid xenon temperature, the leakage current is about 1 nA.  Fig.~\ref{fig:ManyQupid_Breakdown_Temperature_Dependence} shows the temperature dependence of the breakdown voltage for various \textsc{Qupid}s. At -100$^{\circ}$~C the breakdown voltage is about 180 V.  The total gain of the \textsc{Qupid} includes the bombardment gain, given by the photoelectrons impinging onto the APD, and the avalanche gain, given by the avalanche process inside the APD.\\
\indent To measure the gain we have used a pulsed laser light of 405 nm wavelength and $70\pm 30$ ps pulse width emitted at a rate of 100 kHz.\footnote{Hamamatsu Model C10196 Laser Controller with Model M10306-30 PLP-10 Laser Head.} The photocathode voltage\footnote{Stanford Research Systems Model PS355 -10 kV Power Supply} was set to -6 kV and the APD bias voltage just below the breakdown. In order to minimize the cabling passing through the cryostat, both the signal and the bias voltage were carried by the same coaxial cable. To read out the signal, a decoupling circuit was placed outside of the cryostat, along with an amplifier\footnote{RFBay Model LNA-1440 Amplifier, 40 dB gain, 1.4 GHz bandwidth} and an oscilloscope for data acquisition.\footnote{LeCroy WaveRunner 204MXi-L Oscilloscope, 10 GS/s, 2 GHz bandwidth} Fig.~\ref{fig:GainReadout} shows the schematic of the readout used for the gain measurements.\\
\indent To measure the bombardment gain of the \textsc{Qupid}, we integrated the signals read out on the oscilloscope for different photocathode voltages, up to -6 kV. We proceeded as follows:
\begin{itemize}
\item[({\it i})] the gain values obtained are in arbitrary units and must be decoupled from the avalanche gain;
\item[({\it ii})] at large voltages (where the curve shows a linear trend) for each 3.6 eV from the impinging photoelectron, one electron-hole pair is created in the APD~\cite{Silicon};
\item[({\it iii})] the gain curve is then scaled in order to set an increase of one unit in the gain for every 3.6 V.
\end{itemize}
The measurement results of this procedure is shown in Fig.~\ref{fig:BA0011_BombardmentGain}, where the bombardment gain versus the photocathode voltage is reported for different temperatures. The nonlinear trend at negative voltages smaller than -4 kV is due to a dead layer of the APD, which stops the photoelectrons that do not have enough energy to penetrate into the active area.  At -6 kV, the \textsc{Qupid} achieves a bombardment gain of above 700, independent of the temperature.\\
\indent The avalanche gain is measured in a similar fashion, scanning in this case the APD bias voltage just below the breakdown value, keeping the photocathode voltage at -6 kV. To obtain the absolute normalization we:
\begin{itemize}
\item[({\it i})] determine the total gain from a single photoelectron spectrum;
\item[({\it ii})] divide this value by the previously obtained bombardment gain, thereby extracting the avalanche gain; 
\item[({\it iii})] scale the gain versus bias voltage curve to make it to match the value obtained in ({\it ii}) for that specific bias voltage.
\end{itemize}
In Fig.~\ref{fig:BA0011_AvalancheGain} the avalanche gain value versus the bias voltage is reported.  A strong temperature dependence of the avalanche gain can be seen, with the gain increasing as the temperature decreases.\\
\indent The total gain of the \textsc{Qupid} is then $700 \times 200 \sim 10^5$, enough for single photoelectron detection.
\begin{figure*}
  \begin{center}
    \includegraphics[ height=65mm ]{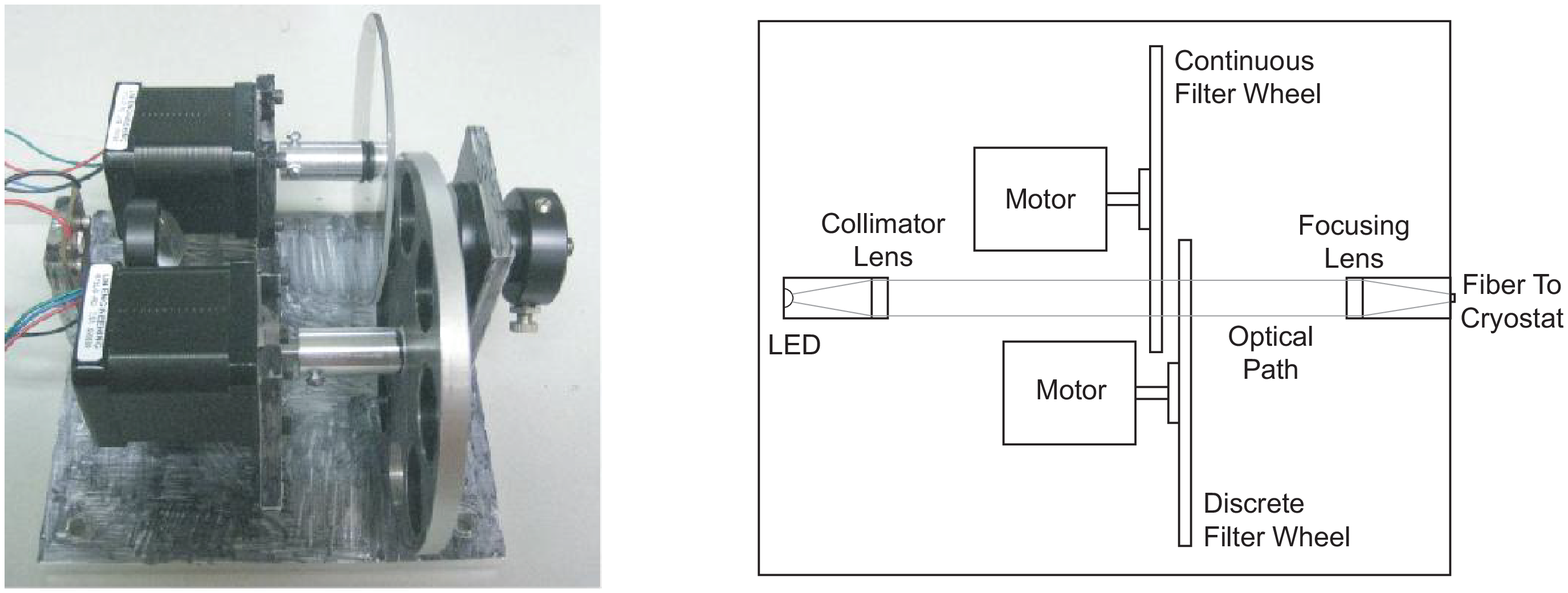}
  \end{center}
  \caption{
    \label{fig:LinearityPulser} 
    On the {\it left}, a photograph of the anode linearity pulsing system. On the {\it right}, a diagram of the anode linearity pulsing system. A pulsed LED shines through two sets of filters and is passed into a fiber, which then carries the light to the \textsc{Qupid}.}
\end{figure*}
\begin{figure}
  \begin{center}
    \includegraphics[ height=65mm ]{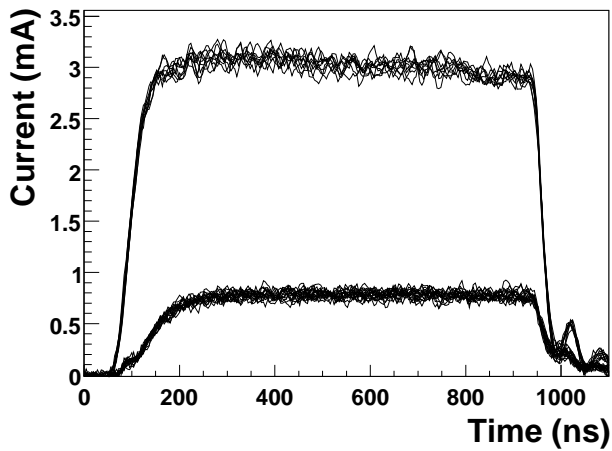}
  \end{center}
  \caption{
    \label{fig:BA0011_Linearity_Waveforms} 
    Waveforms from \textsc{Qupid} (No.5) at the bright and dim light levels from the LED of the linearity testing system. The waveforms from the bright pulses are at 3 mA, where the \textsc{Qupid} starts showing a deviation from linearity at the 5\% level.}
\end{figure}
\begin{figure}
\begin{center}
\includegraphics[ height=65mm ]{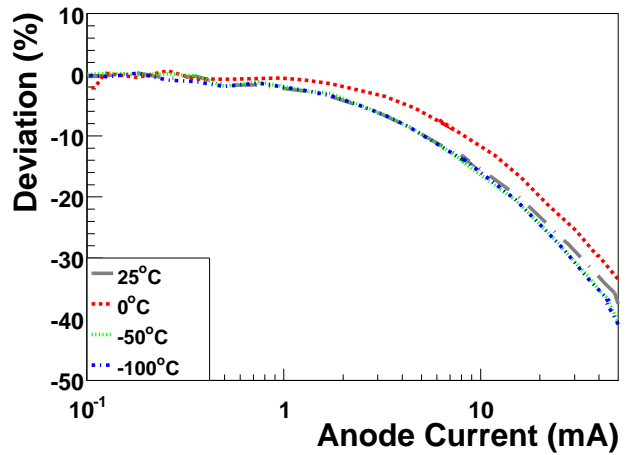}
\end{center}
\caption{
  \label{fig:BA0010_anode_linearity} 
  Anode linearity for various temperatures of \textsc{Qupid} (No.5). A 5\% nonlinear behavior starting at an anode current of 3 mA is evident.  No temperature dependence is observed.}
\end{figure}
\begin{figure*}
  \begin{center}
    \includegraphics[ height=75mm ]{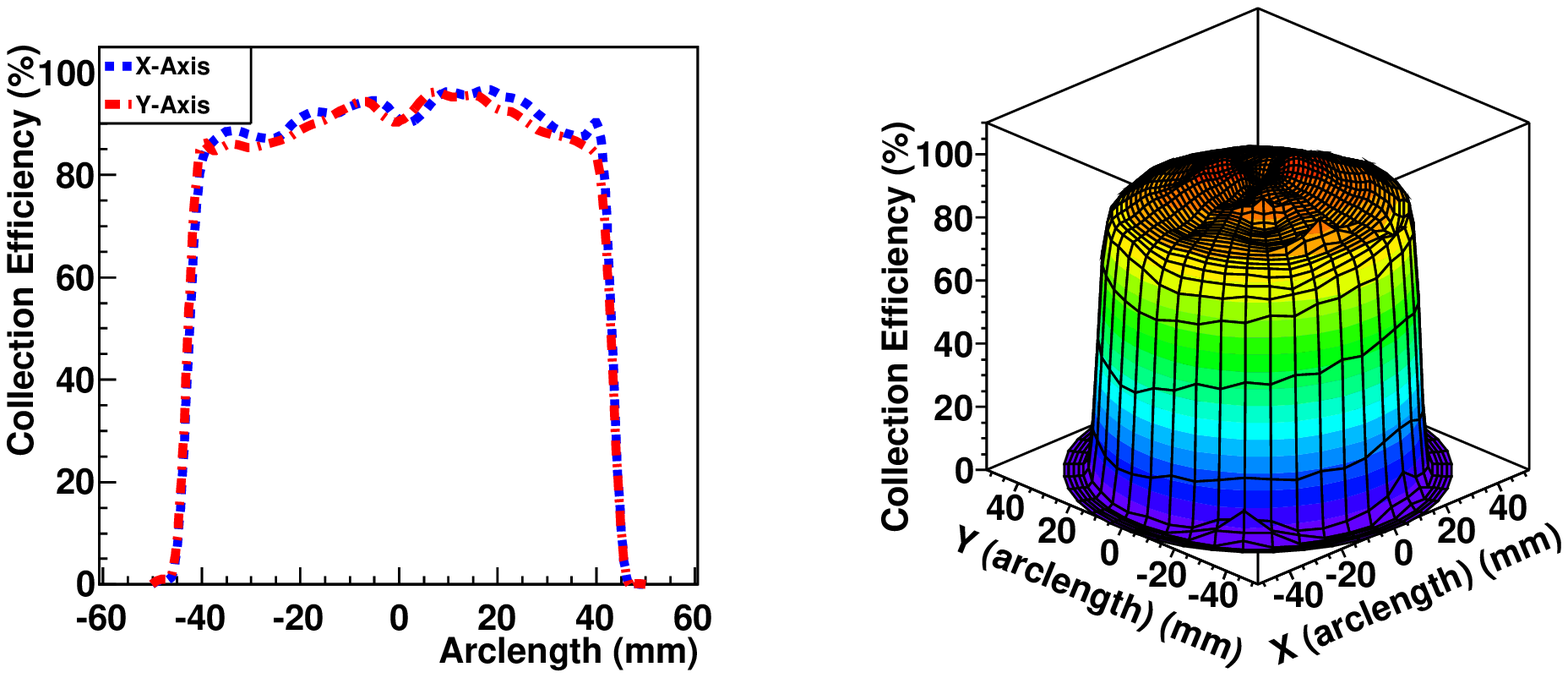}
  \end{center}
  \caption{
    \label{fig:CollectionEff}
    Photoelectron collection efficiency in X and Y slices (on the {\it left}), and in a 3D plot (on the {\it right}) for \textsc{Qupid} (No.7). The collection efficiency is $>80\%$ across the entire face of the \textsc{Qupid}.}
\end{figure*}
\subsection{Linearity}
\label{subsec:anodelinearity}

As discussed in Sec.~\ref{subsec:cathodelinearity}, the photocathode must be linear up to at least 12~nA. The anode must then be linear up to this value times the gain of the \textsc{Qupid}.  For a gain of $10^5$, the \textsc{Qupid} anode should be linear to 1.2~mA.\\
\indent The setup for the linearity characterization of the \textsc{Qupid} consists of two rotating neutral density filter wheels controlled by two stepper motors. One is made of a continuously variable filter wheel with a range of optical densities from 0 to 2.0, the other is a set of discrete filters with optical densities ranging from 0 to 5.0.\footnote{The optical density $A$ is defined as $A=\log_{10}(I_{0}/I)$, where $I_0$ and $I$ are the intensity of the incoming and outgoing lights respectively.} An ultrabright LED\footnote{Nichia Model NS6B083T 470 nm LED} provides pulsed light which is attenuated through the filters and is brought to the \textsc{Qupid} through an optical fiber. Fig.~\ref{fig:LinearityPulser} shows a photograph and a diagram of the testing setup.\\
\indent For the measurement, the LED is pulsed at alternating light levels with a fixed brightness ratio of 1:4, resulting in alternating high and low currents from the APD. Each pulse has a width of 1 $\mu$s, and the dim and bright pulses alternate at a frequency of 600 Hz. The light from the LED is then attenuated through the two filter wheels. In this way, while the overal brightness is changed using different combinations of the filters, the 1:4 brightness ratio is maintained. The waveforms are then read out with an oscilloscope and integrated.\\
\indent At low light levels for both dim and bright pulses, the anode current of the \textsc{Qupid} shows the same constant ratio of 1:4. As the light level increases (by decreasing the optical density of the filter wheels) the absolute value of the anode current increases, and the ratio of low and high anode current starts to deviate from the original 1:4, thereby showing nonlinear behavior. Fig.~\ref{fig:BA0011_Linearity_Waveforms} shows the output current from the \textsc{Qupid} for the bright and dim light pulses while Fig.~\ref{fig:BA0010_anode_linearity} shows the measured anode linearity.\\
\indent Although nonlinear behavior begins to appear at the 5\% level from 3 mA peak anode current with a gain of 10$^{5}$, the nonlinearity of the \textsc{Qupid} can be characterized and corrected for high light levels, thereby increasing the effective dynamic range.\\
\indent Tests at -100$^{\circ}$~C in the cryogenic setup described above prove that the linearity of the \textsc{Qupid} is independent of the temperature.

\subsection{Collection Efficiency}
\label{subsec:anodeuniformity}

The same scanning system used to test the photocathode uniformity (see Sec.~\ref{subsec:cathodeuniformity} and Fig.~\ref{fig:UniformitySystem}) has also been used for  the anode uniformity. In these measurements, the photocathode was held at -6 kV while a bias voltage of -250~V was applied to the anode of the APD. The current was read through the cathode of the APD using a picoammeter.  In this case, the bias on the APD was kept low since the LED used provided a very bright light.  This way, any saturation or nonlinearity of the APD can be avoided.
\begin{figure}
  \begin{center}
    \includegraphics[ height=65mm ]{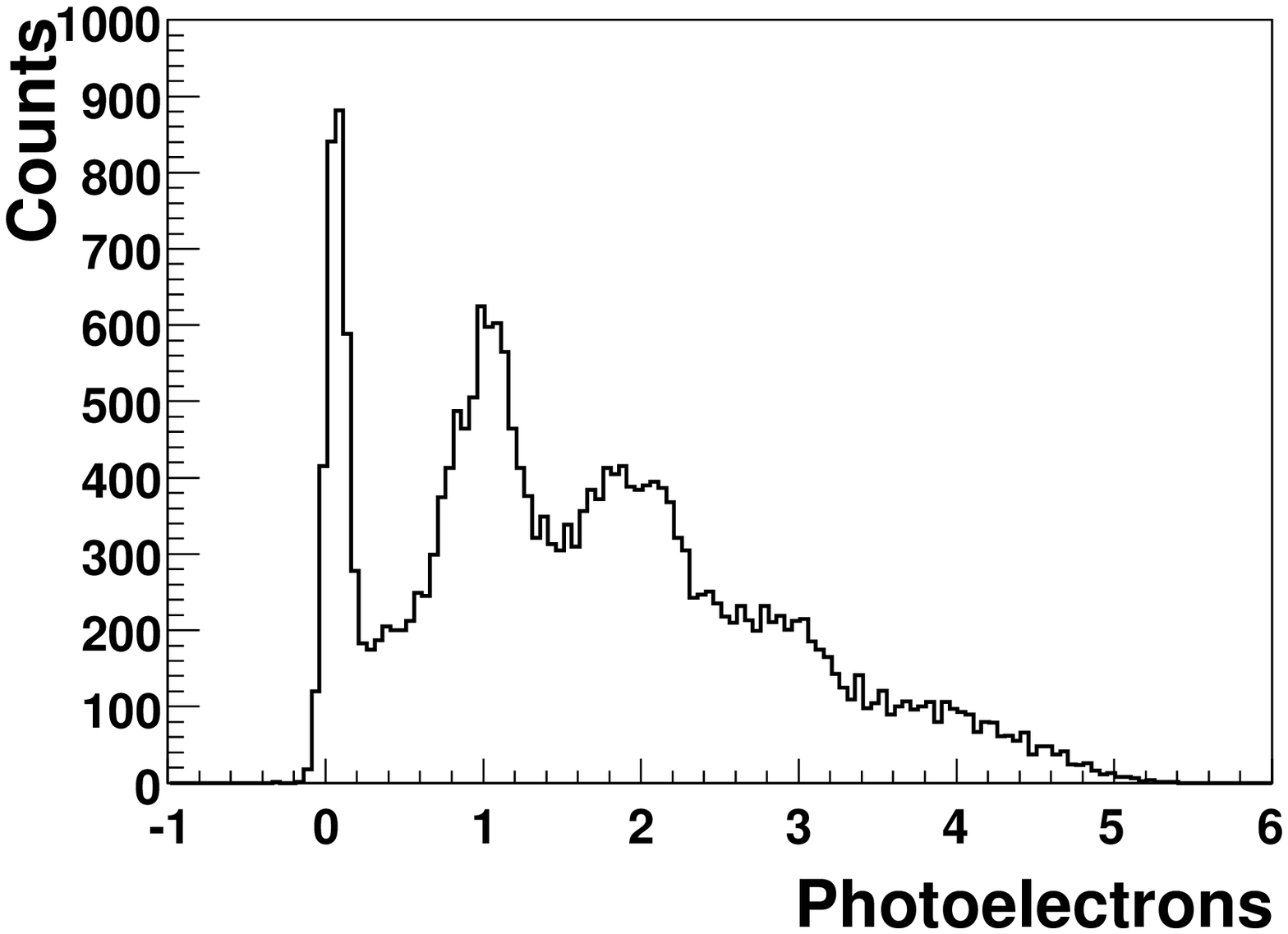}
  \end{center}
  \caption{
    \label{fig:ChargeHist} 
    Charge distribution measured for dim laser pulses on \textsc{Qupid} (No.5).  Peaks of 0, 1, 2, and 3 photoelectrons can be clearly seen.  A narrow pedestal of width 0.09 photoelectrons is visible.}
\end{figure}
\begin{figure}
\begin{center}
  \includegraphics[ height=65mm ]{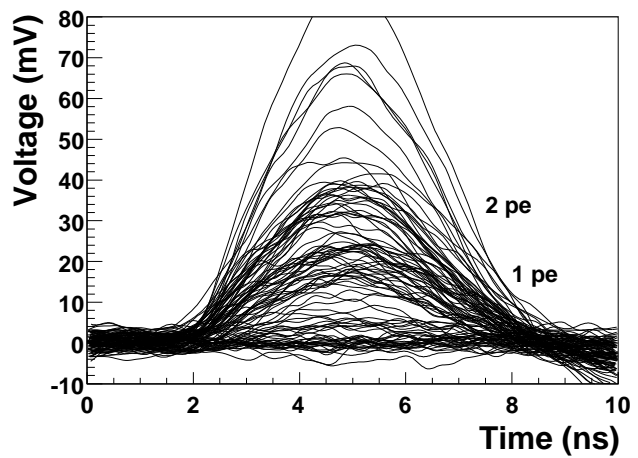}
\end{center}
\caption{
  \label{fig:BA0010_Waveforms} 
  Waveforms for dim laser pulses at -100$^{\circ}$~C from \textsc{Qupid} (No.5).  A 4 m coaxial cable was used between the cryostat and the decoupling circuit. Even with this long cable, clear bands corresponding to 0, 1, and 2 photoelectrons are well visible. The rise time and fall time were measured as $1.8 \pm 0.1$ ns and $2.5\pm 0.2$ ns respectively, with a pulse width of $4.20\pm 0.05$ ns.}
\end{figure}
\begin{figure*}
  \begin{center}
    \includegraphics[ width=160mm ]{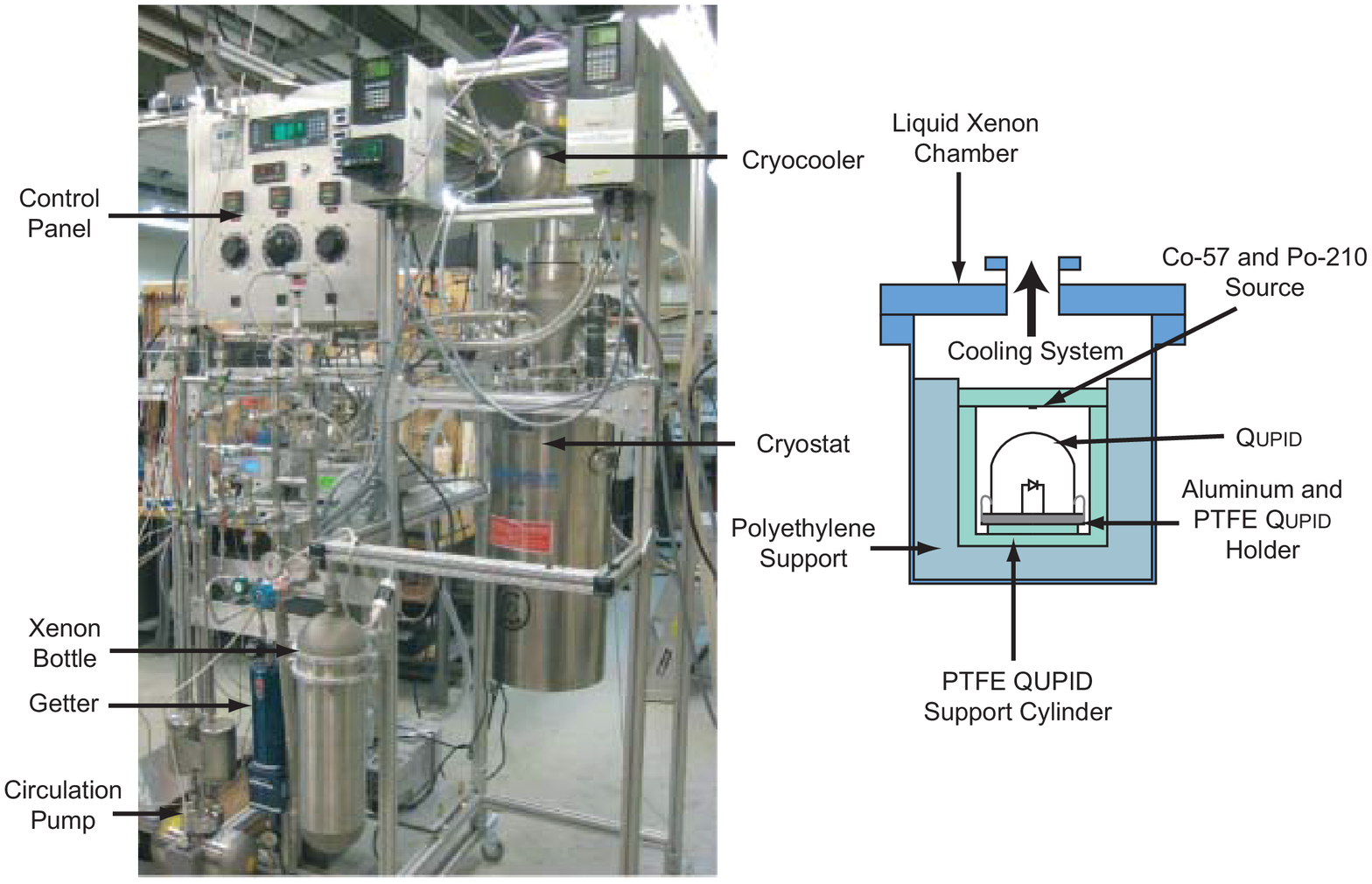}
  \end{center}
  \caption{
    \label{fig:XenonSystem} 
    On the {\it left}, photograph of the liquid xenon test system. On the {\it right}, drawing of the liquid xenon cell inside the cryostat. The \textsc{Qupid} is held in a polyethylene, PTFE, and aluminum holder in liquid xenon, and two radioactive sources are placed inside to generate scintillation light.}
\end{figure*}
The photoelectron collection efficiency can be inferred from the ratio between anode uniformity and photocathode uniformity scaled by the total gain. Fig.~\ref{fig:CollectionEff} shows the photoelectron collection efficiency to be above $80\%$ for a majority of the surface.

\section{Waveforms and Timing}
\label{sec:shapingtiming}

In conjunction with the gain measurements, waveforms of the \textsc{Qupid} were also obtained at low light levels. With the \textsc{Qupid} in the same setup for gain measurements, the intensity of the picosecond laser was lowered such that only a small number of photoelectrons were observed. The picosecond laser controller provided a trigger output synchronized to the laser pulse. A very narrow window was then chosen around this point for integration of the signal, and the resulting charge of the signals were plotted in a histogram. Fig.~\ref{fig:ChargeHist} shows the charge distribution of \textsc{Qupid} signals from 0, 1, 2, and 3 photoelectrons. A narrow pedestal of zero photoelectrons can be seen in the histogram, and clear peaks corresponding to integer number of photoelectrons are apparent.\\
\indent Since ton-scale dark matter detectors will require long cabling of the order of several meters, to ensure that the readout electronics are far from the target volume, we included a 4 m coaxial cable between the cryostat and the decoupling circuit. Fig.~\ref{fig:BA0010_Waveforms} shows 100 waveforms registered at -100$^{\circ}$~C. Even with such a long cable before amplifying the signal, there is little degradation of the waveform, and bands of 0, 1, and 2 photoelectrons can be clearly seen.\\
\indent Timing information of the waveforms was measured in the same setup. Defining the rise (fall) time as the time for the pulse to go from 10\% to 90\% of the pulse height (and vice-versa), we obtained a rise time of $1.8\pm 0.1$ ns and a fall time of $2.5\pm 0.2$ ns. The pulse width, defined as the full width at half maximum (FWHM) of the waveforms, is $4.20\pm 0.05$ ns.  The transit time spread of the \textsc{Qupid} has been measured by comparing the time difference between the laser trigger and the peak of the waveform. Histogramming this time difference and determining the FWHM, we obtained $160\pm 30$~ps. It is important to note that this value can be considered as an upper limit as it includes the uncertainties arising from the jitter of the laser trigger.
\begin{figure*}
  \begin{center}
    \includegraphics[trim = 1cm 0cm 1cm 0cm, clip, height=115mm ]{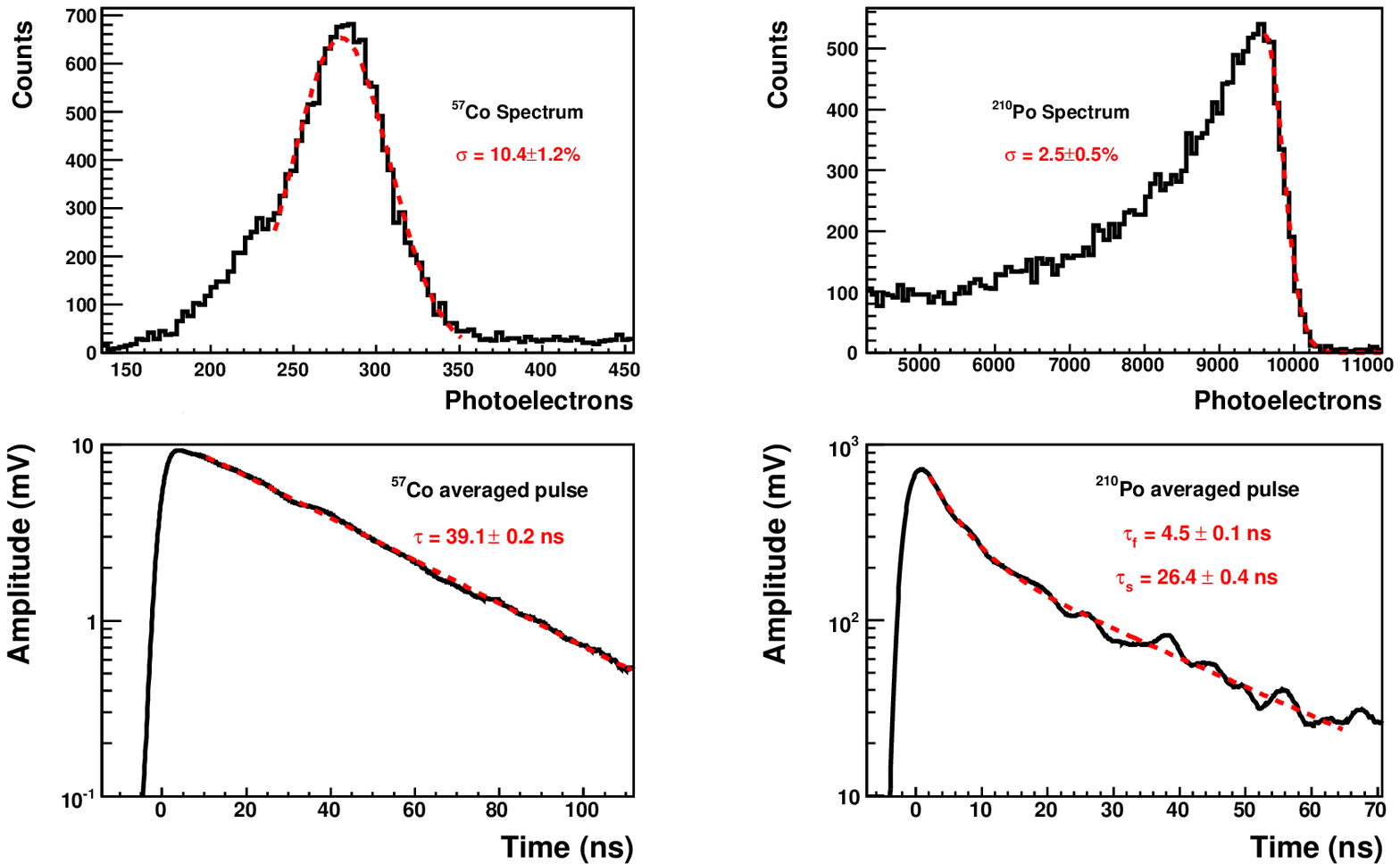}
\end{center}
\caption{
  \label{fig:4plotpaperlog} 
  Clockwise from top left: $^{57}$Co energy spectrum, $^{210}$Po energy spectrum, $^{210}$Po averaged waveform, $^{57}$Co averaged waveform using \textsc{Qupid} (No.7). The average light yield from several data sets of $^{57}$Co is $2.0\pm 0.2$ photoelectrons/keV (pe/keV) with a resolution of $10.4 \pm 1.2\%$.  The average light yield obtained from the $^{210}$Po source was $1.6\pm 0.2$ pe/keV with a resolution of $2.5\pm 0.5\%$. The average waveforms also show exponential fits in red. A decay time of $39.1\pm 0.2$ ns is found from the $^{57}$Co pulses, and a fast and slow decay time of $4.5\pm 0.1$ ns and $26.4\pm 0.4$ ns are seen in the $^{210}$Po waveforms.}
\end{figure*}
\section{\textsc{Qupid} in liquid xenon}
\label{sec:liquidxenon}

\begin{table*}
  \centering
  \renewcommand{\arraystretch}{1.2}
  \begin{tabular}{|l|c|c|c|c|c|c|}
    \hline
    \multirow{2}{*}{Source} & \multirow{2}{*}{Type} & \multirow{2}{*}{Energy} & \multirow{2}{*}{Light Yield} & \multirow{2}{*}{Resolution} & \multirow{2}{*}{Decay Time} & Previously Measured\\
    & & & & & &  Decay Time~\cite{Hitachi,Kubota}\\
    \hline
    \hline
    \multirow{2}{*}{$^{57}$Co} & \multirow{2}{*}{$\gamma$} & 122 keV, 86\% & \multirow{2}{*}{$2.0\pm 0.2$ pe/keV} & \multirow{2}{*}{$10.4\pm 1.2$\%} & \multirow{2}{*}{$39.1\pm 0.2$ ns} & \multirow{2}{*}{$34\pm 2$ ns}\\
    & & 136 keV, 11\% & & & & \\
    \hline
    \multirow{2}{*}{$^{210}$Po} & \multirow{2}{*}{$\alpha$} & \multirow{2}{*}{5.3 MeV} & \multirow{2}{*}{$1.6\pm 0.2$ pe/keV} & \multirow{2}{*}{$2.5\pm 0.5$\%} & $4.5\pm 0.1$ ns, fast (71\%) & $4.3\pm 0.6$ ns, fast (69\%)\\
    & & & & & $26.4\pm 0.4$ ns, slow (29\%) & $22.0\pm 2.0$ ns, slow (31\%)\\
    \hline
  \end{tabular}
  \caption{Parameters for the sources observed in liquid xenon, including light yield, resolution, and decay times.  This particular \textsc{Qupid} had a lower quantum efficiency of 20\% at 178 nm.  }
  \label{tab:LXeSources}
\end{table*}
\begin{table*}
  \centering
  \setlength{\extrarowheight}{1.5pt}
  \setlength{\tabcolsep}{20pt}
  \renewcommand{\thefootnote}{\thempfootnote}
  \begin{tabular}{|@{}l|@{}l|c c|}
    \hline
    \multicolumn{4}{|c|}{\textsc{Qupid} Parameters} \\
    \hline
    \hline
    \multirow{3}{*}{Dimensions} & Outer Diameter & \multicolumn{2}{c|}{71 mm} \\
    & Effective Photocathode Diameter & \multicolumn{2}{c|}{64 mm} \\
    & Radius of Hemispherical Photocathode & \multicolumn{2}{c|}{37 mm}\\
    & Total Height & \multicolumn{2}{c|}{76 mm} \\
    \hline
    \multirow{5}{*}{Radioactivity} & $^{238}$U & \multicolumn{2}{c|}{$<17.3$ mBq}\\
    & $^{226}$Ra & \multicolumn{2}{c|}{$0.3\pm 0.1$ mBq}\\
    & $^{232}$Th & \multicolumn{2}{c|}{$0.4\pm 0.2$ mBq}\\
    & $^{40}$K & \multicolumn{2}{c|}{$5.5\pm 0.6$ mBq}\\
    & $^{60}$Co & \multicolumn{2}{c|}{$<0.18$ mBq}\\
    \hline
    \multicolumn{2}{|c|}{Performance} & 25$^{\circ}$ C & -100$^{\circ}$~C\\
    \hline
    \multirow{3}{*}{Photocathode} & Material & \multicolumn{2}{c|}{Bialkali-LT}\\
    & Quantum Efficiency at 178 nm & $34\pm 2$\% & -- \\
    & Linearity\footnote{Maximum current within 5\% nonlinearity} & $>10$ $\mu$A & $>1$ $\mu$A \\
    \hline
    \multirow{3}{*}{Electron Bombardment} & Acceleration Voltage & \multicolumn{2}{c|}{6 kV}\\
    & Typical Gain &  \multicolumn{2}{c|}{750} \\
    & Maximum Gain &  \multicolumn{2}{c|}{800} \\
    \hline
    \multirow{6}{*}{APD} & Diameter & \multicolumn{2}{c|}{3 mm}\\
    & Capacitance & \multicolumn{2}{c|}{11 pF}\\
    & Leakage Current & 200 nA & 0.3 nA \\
    & Breakdown Voltage & 360 V & 180 V \\
    & Typical Gain &  \multicolumn{2}{c|}{200} \\
    & Maximum Gain &  \multicolumn{2}{c|}{300} \\
    \hline
    \multirow{3}{*}{Anode Output} & Typical Total Gain &  \multicolumn{2}{c|}{$1.5\times 10^{5}$}\\
    & Maximum Total Gain &  \multicolumn{2}{c|}{$2.4\times 10^{5}$}\\
    & Linearity\footnotemark[\value{mpfootnote}] &  \multicolumn{2}{c|}{3 mA}\\
    \hline
    \multirow{4}{*}{Timing Properties} & Rise Time (10\%-90\%) &  \multicolumn{2}{c|}{$1.8\pm 0.1$ ns}\\
    & Fall Time (90\%-10\%) &\multicolumn{2}{c|}{$2.5\pm 0.2$ ns}\\
    & Pulse Width (50\%-50\%)  & \multicolumn{2}{c|}{ $4.20\pm 0.05$ ns}\\
    & Transit Time Spread (FWHM) &  \multicolumn{2}{c|}{$160\pm 30$ ps}\\
    \hline
  \end{tabular}
  \caption{Summary of the key parameters of the \textsc{Qupid}.}
  \label{tab:Summary}
\end{table*}

The \textsc{Qupid} has been tested extensively in a liquid xenon setup built at UCLA, which is shown in Fig.~\ref{fig:XenonSystem}. The \textsc{Qupid} was placed inside a PTFE and aluminum housing supported by a polyethylene structure inside a stainless steel chamber. The chamber was then placed inside a vacuum cryostat. A cryocooler\footnote{Q-Drive Model 2S132K-WR Cryocooler} was used to liquefy the xenon and to maintain liquid xenon temperature during operation. The PTFE housing was fully immersed in the liquid xenon allowing for the \textsc{Qupid} to operate in single-phase mode. The temperature and pressure were held constant at -100$^{\circ}$~C and 1.6 bar, and the system was operated under stable conditions for approximately two weeks. During this period, the xenon gas was purified with a hot metal getter\footnote{SAES Model PS3MT3R1 Mono-torr Getter} in a closed recirculation loop\footnote{Q-Drive/UCLA Model 2S132K-UCLA Pump}.\\
\indent Gain calibration of the \textsc{Qupid}  was performed by measuring its response to light from the picosecond laser, fed into the PTFE housing via an optical fiber.  The readout setup was the same as for the gain measurements (Fig.~\ref{fig:GainReadout}), however no amplifier was used.  Internal $^{57}$Co and $^{210}$Po sources,  placed just above the \textsc{Qupid}, were used to measure and monitor the \textsc{Qupid} response to the scintillation light of xenon.  The presence of impurities, out-gassing of the surrounding material, and a PTFE housing that was not fully optimized limited the achievable light detection efficiency, which was found to be strongly dependent on the recirculation speed throughout the run. Fig.~\ref{fig:4plotpaperlog} shows the response of the \textsc{Qupid} to the $^{57}$Co and $^{210}$Po sources, including pulse shapes and energy spectra.\\
\indent Table~\ref{tab:LXeSources} summarizes the parameters measured from each of the sources. $^{57}$Co decays through the emission of a 122 keV $\gamma$-ray with an 86\% branching ratio, and a 136 keV $\gamma$-ray with 11\% branching ratio.  As it can be seen in Fig.~\ref{fig:4plotpaperlog}, these two lines are not separated but give a single peak whose weighted average energy of 123.6 keV was used for further calculations.  The $^{57}$Co light yield averaged over a few hours was found to be $2.0\pm 0.2$ pe/keV, with $10.4 \pm 1.2\%$ energy resolution. The stated errors are estimated from the fluctuations of the light yield for the measurements taken with different trigger threshold values over the span of a few hours.\\
\indent Fig.~\ref{fig:4plotpaperlog} also shows the part of the measured spectrum dominated by the peak due to the $^{210}$Po source. $^{210}$Po decays by a 5.3 MeV $\alpha$-particle, and from this a light yield of $1.6 \pm 0.2$ pe/keV was obtained with a resolution of $2.5 \pm 0.5\%$.  These measurements were taken on a different date than the $^{57}$Co data, and thus the conditions, such as xenon purity, changed between the data sets.\\
\indent The scintillation light from liquid xenon has two decay components due to the de-excitation of the singlet and triplet states of the excited dimer Xe$^{\ast}_{2}$~\cite{Scint}. The averaged pulse shape for $^{210}$Po $\alpha$-particle and $^{57}$Co $\gamma$-ray interactions, together with the fit of their overall decay profiles, are shown in Fig.~\ref{fig:4plotpaperlog} (bottom).  The $\alpha$-particle interactions from $^{210}$Po had measured fast and slow decay times of $4.5 \pm 0.1$ ns and $26.4 \pm 0.4$ ns respectively, with an intensity ratio of the fast to slow components being 71\% fast and 29\% slow.  The $^{57}$Co pulses had a single decay time of $39.1\pm 0.2$ ns.  These decay time constants are similar to the previously published values of $4.3 \pm 0.6$ ns and $22.0 \pm 2.0$ ns for the $\alpha$-particle fast and slow components, with a ratio of 69\% fast and 31\% slow, and to $34\pm 2$ ns for $\gamma$-interactions~\cite{Hitachi,Kubota}.  The ability to observe a difference in the pulse shapes of $\alpha$-particle and $\gamma$-ray interactions allows for the possibility of pulse shape discrimination in dark matter detectors.  Although the number of photoelectrons expected in a liquid xenon dark matter detector is too low for such discrimination, the technique can be applied to liquid argon detectors~\cite{PSD1,PSD2,PSD3,PSD4}.

\section{Conclusion}
\label{sec:conclusions}

The \textsc{Qupid}s, the new low radioactivity photodetectors developed and evaluated jointly by Hamamatsu Photonics and UCLA, show optimum characteristics for use in the next generation of dark matter and double beta decay detectors.  In this work we have described the test setups and performance of seven \textsc{Qupid}s. Table~\ref{tab:Summary} shows a summary of the most relevant results.\\
\indent\textsc{Qupid}s have much lower radioactivity than conventional PMTs as measured at the Gator screening facility. Simulations of ton scale detectors including \textsc{Qupid}s show that they satisfy the low background level requirement.  The quantum efficiency, higher than 30\% at the xenon scintillation wavelength, is competitive with standard photodetectors. A total gain of 10$^{5}$, and the capability of single photon counting, fit very well with the detection of low intensity signals coming from WIMP-nuclei interactions. A wide linear dynamic range, up to 3 mA of anode current at liquid xenon temperature, allows the \textsc{Qupid} to cover an energy range large enough for neutrinoless double beta decay detection.	We have measured the \textsc{Qupid} uniformity to be above 80\% over the entire surface, and have measured a good timing response of $1.8 \pm 0.1$~ns rise time, $2.5 \pm 0.2$~ns fall time, $4.20 \pm 0.05$~ns (FWHM) pulse width, and $160\pm 30$~ps (FWHM) transit time spread. In a liquid xenon environment, we have demonstrated the \textsc{Qupid}s capability to detect scintillation light from $\gamma$-ray and $\alpha$-particle interactions.\\
\indent These characteristics make the \textsc{Qupid} an ideal replacement for PMTs in future experiments, such as \textsc{DarkSide}50, XENON1Ton, MAX, DARWIN, and XAX~\cite{DarkSide,XAX,Xenon1T,DARWIN,MAX}, and will represent a major contribution for the next generation of ton scale dark matter and double beta decay detectors.

\section{Acknowledgements}
We would like to thank J. Takeuchi, T. Hakamata, and Y. Hotta for their in-kind contributions and management through the \textsc{Qupid} development and manufacturing at Hamamatsu Photonics, and Y. Egawa, A. Kageyama, Y. Negi, and M. Yamada for the final design of the \textsc{Qupid} along with the development of the \textsc{Qupid} manufacturing line and actual manufacturing.  We would also like to thank Y. Ishikawa and K. Yamamoto from the Solid State division of Hamamatsu Photonics for the development of a new APD specifically optimized for the \textsc{Qupid}.  Special thanks to D. Dreisbach, P. Hemphill, M. Levi, C. Reilly, J. Rolla, and T. White for their invaluable help in the development of the \textsc{Qupid} test systems, data taking and analysis, and H. Lockhart and the UCLA Machine Shop for the machining of many of the parts used in the test setups.  We are indebted to E. Aprile, F. Calaprice, C. Galbiati, B. Sadoulet, and M. Tripathi for useful discussions. We also express our deep gratitude to the XENON collaboration for helpful advice and encouragement.  This work was supported in part by US DOE grant DE-FG02-91ER40662, and by NSF grants PHY-0653459, PHY-0810283, PHY-0919363, and PHY-0904224, and we gratefully thank H. Nicholson and M. Salamon from the DOE, and J. Whitmore, J. Kotcher, and D. Lissauer from the NSF.  Additional support was provided by the INPAC Fund from the UC Office of the President, the UCLA Dean, and UCLA Physics Chair funds, and we are thankful to R. Peccei, J. Rudnick, and F. Coroniti for financial support and encouragement.  The work done at the Gator facility was supported by the Swiss National Foundation, Grant No. 20-118119 and No. 20-126993.

\bibliographystyle{model1-num-names}

\end{document}